\documentclass[twocolumn,floatfix,prb,aps,showpacs,superscriptaddress,longbib,longbibliography]{revtex4-2} 
\bibpunct{[}{]}{,}{n}{}{}

\usepackage{graphicx,epsf,amsmath,amssymb,amsfonts}
\usepackage{bm,bbm}
\usepackage[usenames,dvipsnames,svgnames,table]{xcolor}
\usepackage[colorlinks,citecolor=blue,urlcolor=blue]{hyperref}
\usepackage{empheq}
\usepackage{ulem}
\usepackage{color}
\usepackage[b]{esvect} 
\usepackage[colorlinks,citecolor=blue,urlcolor=blue]{hyperref}
\usepackage{float}
\usepackage{cases}
\usepackage{fancyhdr}
\hypersetup{colorlinks=true}
\usepackage[all]{hypcap}

\begin{document}

\title{Bootstrapping Bloch bands}

\author{Serguei Tchoumakov} 
\affiliation{Univ. Grenoble Alpes, CNRS, Grenoble INP, Institut N\'eel, 38000 Grenoble, France}

\author{Serge Florens} 
\affiliation{Univ. Grenoble Alpes, CNRS, Grenoble INP, Institut N\'eel, 38000 Grenoble, France}

\date{\today}

\begin{abstract}
    Bootstrap methods, initially developed for solving statistical and quantum field theories, have recently been shown to capture the discrete spectrum of quantum mechanical problems, such as the single particle Schrödinger equation with an anharmonic potential. The core of bootstrap methods builds on exact recursion relations of arbitrary moments of some quantum operator and the use of an adequate set of positivity criteria. We extend this methodology to models with continuous Bloch band spectra, by considering a single quantum particle in a periodic cosine potential. We find that the band structure can be obtained accurately provided the bootstrap uses moments involving both position and momentum variables. We also introduce several new techniques that can apply generally to other bootstrap studies. First, we devise a trick to reduce by one unit the dimensionality of the search space for the variables parametrizing the bootstrap. Second, we employ statistical techniques to reconstruct the distribution probability allowing to compute observables that are analytic functions of the canonical variables. This method is used to extract the Bloch momentum, a quantity that is not readily available from the bootstrap recursion itself.
\end{abstract}

\maketitle

\section{Introduction}
Despite the paradoxical name, bootstrapping methods constitute an interesting
toolkit to solve a large variety of theoretical physics models. The essence of this
idea lies in setting an infinite set of internal constraints that can be derived from
the studied problem, and that can be used to determine uniquely its solution {\it without ever
computing it explicitly}. While the initial bootstrap program is aimed at constructing
fundamental quantum field theories allowed by Nature~\cite{bootstrap_Smatrix}, later developments focused on 
practical solutions in statistical and quantum mechanics~\cite{bootstrap_CFT}. 
Quite remarkably, enforcing conformal invariance helps formulate a bootstrap principle to compute the critical 
exponents of the three-dimensional (3D) Ising model~\cite{bootstrap_Ising} and solve some of its correlations 
functions~\cite{bootstrap_corr}, while not feasible with other techniques.
More recently, a class of many-body quantum matrix models are shown to be amenable to a bootstrap solution~\cite{bootstrap_matrix}, from which the single particle Schrödinger equation with quartic potential is a particular case~\cite{bootstrap_qm}. Here, the bootstrap method relies on the relationship between the statistical moments of quantum operators, related to the positivity of a quadratic form, and that is used to check whether the recursion between moments of an eigensolution of the Hamiltonian is physical or not. This method has already been implemented to various models in quantum mechanics \cite{bhattacharya2021numerical,berenstein2021bootstrapping,berenstein2021bootstrapping2,aikawa2021application,aikawa2021comment}.
%

Our main goal here is to extend the bootstrap methodology to solve periodic quantum Hamiltonians
with continuous spectra and to propose new tools to improve the convergence and application-range of the bootstrap.
We study in detail the solution to a single quantum particle in a periodic cosine potential.
We show that the convergence of the bootstrap depends crucially on the appropriate choice of
constraints. Indeed, we find that the use of statistical moments that mix
position and momentum drastically improves the convergence with respect to a bootstrap involving only the moments of a single canonical variable. The band structure of the cosine potential problem is well reproduced by the bootstrap, and convergence seems uniform. Due to the amplification of numerical round-off errors in the bootstrap recursion, arbitrary precision numerics need to be used in order to reach machine precision for the band spectrum.

In previous bootstrap studies~\cite{bootstrap_Ising,bootstrap_corr,bootstrap_matrix,bootstrap_qm}, the energy spectra are obtained by scanning a single moment for each energy eigenvalue to find values that satisfy the bootstrap constraints. This exploration over a two-dimensional (2D) space is numerically heavy and we devise a dimensional reduction method that shrinks the search space from 2D to 1D.
Finally, the bootstrap is often used to compute the spectrum and the moments involved in the
recursion relation. The choice of moments is limited to those where the recursion relation can be
closed and thus prevents the evaluation of observables for which no closed recursion relation
exists. For example, we show that the evaluation of the Bloch momentum of the periodic quantum Hamiltonian necessitates to evaluate an infinite recursion. For this type of situation, we propose a probabilistic method, based on the truncated Hamburger moment problem, to compute arbitrary analytic functions of the canonical variables. Our methodology is general enough that it can be extended to other bootstrap studies~\cite{bootstrap_manybody,bootstrap_matrix2,bootstrap_matrix3}.


The paper is organized as follows. In Sec.~\ref{Sec:Bootstrap}, we derive all the useful exact
bootstrap identities for the model of a single quantum particle in a 1D cosine potential. 
In Sec.~\ref{Sec:Solve}, we show a general dimensional reduction of the search space for the bootstrap variables, which can apply to other bootstrap studies. 
In Sec.~\ref{Sec:Distrib}, we construct a distribution that allows to compute the expectation value of an arbitrary analytic function of the canonical variables, illustrating how Bloch's
momentum can be reconstructed. General perspectives close the manuscript.

\section{Bootstrapping identities}
\label{Sec:Bootstrap}
We focus here on the Hamiltonian describing a single quantum particle moving in
a cosine potential on the real line:
\begin{align}
\hat{H} = \hat{p}^2 + 2 v \cos(\hat{x}),
\label{eq:h}
\end{align}
where $[\hat{x},\hat{p}] = i$ and $2v$ is the strength of the potential. 
We will set throughout $\hbar=1$ and the particle mass $2m=1$. 
This simple Hamiltonian is ubiquitous in physics, describing {\it e.g.} the motion of electrons 
in crystals~\cite{Kittel2004}, as well as the dynamics of Josephson junctions~\cite{tinkham2004introduction}.

The relation between averaged observables can be obtained using the following identity~\cite{bootstrap_qm}. In any eigenstate $|\Psi\rangle$ of $\hat{H}$ with energy $E$ and for any operator $\hat{O}$, the following average vanishes
\begin{align}
    \langle [\hat{H}, \hat{O} ] \rangle = \langle\Psi |[\hat{H}, \hat{O} ]|\Psi \rangle 
= \langle\Psi |E \hat{O}-\hat{O}E|\Psi \rangle =0.
    \label{eq:comm}
\end{align}
Due to the periodicity of Hamiltonian~\eqref{eq:h}, we expand~\eqref{eq:comm} for the operators $\hat{O}_1 = \exp({i n \hat{x}})$ and $\hat{O}_2 = \exp(i n \hat{x}) \hat{p}$, with $n$ an integer, leading to the respective identities:
\begin{align}
    n^2 \langle e^{in\hat{x}}\rangle + 2 n \langle e^{in\hat{x}} \hat{p} \rangle = 0,
    \label{eq:b1}
\end{align}
and,
\begin{align}
    n^2\langle e^{in\hat{x}} \hat{p}\rangle + 2 n \langle e^{in\hat{x}} \hat{p}^2\rangle + v(\langle e^{i(n-1)\hat{x}}\rangle - \langle e^{i(n+1)\hat{x}}\rangle)=0.
    \label{eq:b2}
\end{align}
Since the averages are over an eigenstate of $\hat{H}$, we can also use the identity $\langle \hat{O}\hat{H} \rangle = E \langle \hat{O} \rangle$ with $\hat{O} = \exp({i n \hat{x}})$:
\begin{align}
    \langle e^{in\hat{x}} \hat{p}^2\rangle + v(\langle
    e^{i(n-1)\hat{x}}\rangle + \langle e^{i(n+1)\hat{x}}\rangle)
    =E \langle e^{in\hat{x}} \rangle.
    \label{eq:b3}
\end{align}
The identities in Eqs.~(\ref{eq:b1}-\ref{eq:b3}) lead to the closed recursion on the expectation values of the phase operator $\exp({i n \hat{x}})$:
\begin{align}
    \label{eq:SimpleRecursion}
    &n(4E-n^2) \langle e^{in\hat{x}} \rangle \\
    &+ 2v\left((1-2n)\langle e^{i(n-1)\hat{x}} \rangle - (1+2n)\langle e^{i(n+1)\hat{x}} \rangle\right) 
    = 0.\nonumber
\end{align}
One can readily show that $\langle \exp({in\hat{x}}) \rangle = \langle \cos(n\hat{x}) \rangle$, so the recursion relation can be computed for any $n$ once $E$ and $\langle \cos(\hat{x}) \rangle$ are set, since $\langle \exp({i 0 \hat{x} }) \rangle=1$ due to normalization of the probability distribution.
Note that formally the norm $\langle1\rangle$ diverges in the thermodynamic limit due to the non
confining periodic potential, but quantities such as $\langle e^{in\hat{x}} \rangle / \langle 1
\rangle $ should be well defined.

Also, it is useful to derive the recursion between averages in the form $\langle \exp({i n \hat{x} }) \hat{p}^s \rangle$ involving both the phase operator $\exp({i n \hat{x}})$ and an arbitrary power of the momentum operator $\hat{p}$. Following the same approach, we establish the following double recursion (see Appendix~\ref{App} for details):
\begin{eqnarray}
    \nonumber
    &&(n^2 - E)\langle e^{inx} p^{s} \rangle + 2n\langle e^{inx} p^{s+1} \rangle + \langle e^{inx} p^{s+2}
    \rangle \\
    &&+ v(\langle e^{i(n+1)x} p^{s}\rangle + \langle e^{i(n-1)x} p^{s}\rangle ) = 0.
    \label{DoubleRecursion}
\end{eqnarray}
Since $\langle \exp({in\hat{x}}) \rangle$ is known for all $n$ from the single recursion Eq.~\eqref{eq:SimpleRecursion}, and that $\langle \exp({in\hat{x}}) \hat{p}\rangle$ can be deduced from Eq.~(\ref{eq:b1}), we can use Eq.~(\ref{DoubleRecursion}) to compute the moments in the form $\langle \exp({in\hat{x}}) \hat{p}^s\rangle$, for all $s>1$.

\section{Dimensional reduction of the search space}
\label{Sec:Solve}

We now present how the bootstrap conditions are generally derived, before addressing the details of the solution. We first define a general operator $\hat{O} = \sum_{n = 0}^{K} \sum_{s=0}^{L} a_{ns} \hat{p}^s \exp({i n\hat{x}})$ from a set of $(K+1)\times(L+1)$ arbitrary complex coefficients $a_{ns}$. The operator $\hat{O}^{\dagger}\hat{O}$ is necessarily positive semi-definite, so its expectation value in any state satisfies the positivity constraint $\langle \hat{O}^{\dagger}\hat{O} \rangle \geq 0$~\cite{bootstrap_lattice,bootstrap_matrix,bootstrap_qm}. This implies that the matrix $\mathcal{M}_{n\sigma,m\tau} = \langle e^{-imx}\hat{p}^{\sigma + \tau} e^{in\hat{x}} \rangle$ should be positive semi-definite. For averages over an exact eigenstate of $H$ the entries of $\mathcal{M}_{n\sigma,m\tau}$ are obtained from Eq.~(\ref{DoubleRecursion}), see Appendix~\ref{App} for details. Thus, once the energy $E$ and the first moment $\langle \cos(x) \rangle$ are set, the whole matrix $\mathcal{M}$ is determined.

The usual bootstrap consists in scanning all possible values of $E$ and $\langle \cos(x) \rangle$
to check if $\mathcal{M}$ is positive semi-definite, \textit{i.e.} if its eigenvalues are all 
greater or equal to zero. Any pair ($E, \langle \cos(x) \rangle$) that leads to a non-positive $\mathcal{M}$ is non-physical and the region of physical solutions gets more and more precise when increasing $K$ or $L$, \textit{i.e.} when increasing the number of constraint, although the exact wavefunction is never computed explicitly. This procedure is numerically expensive since it requires to scan all possible values of $E$ and $\langle \cos(x) \rangle$ and we now show how the search space can be reduced to the energy $E$ only. 

The statistical moments in Eq.~(\ref{DoubleRecursion}) are linearly related, so the matrix $\mathcal{M}$ evaluated for $(E,\langle \cos(x) \rangle)$ satisfies:
\begin{align}
    \mathcal{M} = \mathcal{M}^{(0)} + \langle \cos(x)\rangle \mathcal{M}^{(1)},
\end{align}
where $\mathcal{M}^{(0)}$ and $\mathcal{M}^{(1)}$ are $\mathcal{M}$ matrices computed for energy $E$ and respectively $\langle \cos(x) \rangle = 0$ and $\langle \cos(x) \rangle = 1$. The eigenvalues $\lambda$ of $\mathcal{M}$ then obey:
\begin{align}
\left( \mathcal{M}^{(0)} + \langle \cos(x) \rangle \mathcal{M}^{(1)} \right) {\bf v} = \lambda {\bf v},
\end{align}
and the regions in the search space $(E,\langle \cos(x) \rangle)$ where $\mathcal{M}$ is definite
positive are delimited by boundaries where at least one of the eigenvalues $\lambda$ vanishes. The value of $\langle \cos(x) \rangle$ at these boundaries thus satisfies the eigenvalue equation:
\begin{align}
[\mathcal{M}^{(1)}]^{-1}\mathcal{M}^{(0)} {\bf v} = - \langle \cos(x) \rangle {\bf v}.
\label{eq:reduction}
\end{align}
The scan of all possible values of $\langle \cos(\hat{x}) \rangle$ is thus reduced to the
$(K+1)\times(L+1)$ discrete eigenvalues of $[\mathcal{M}^{(1)}]^{-1}\mathcal{M}^{(0)}$ for a given choice of $E$. Since $|\langle\cos(\hat{x})\rangle|\leq1$, we can further limit the search to the few eigenvalues that are in the $[-1,1]$ interval.
This technique accelerates solving the bootstrap, which is especially useful to find the continuum of eigenstates in the Bloch Hamiltonian~(\ref{eq:h}). In general, a bootstrap with a $d$-dimensional search space can be reduced to $d-1$ dimensions with this technique.

\begin{figure}[htb]
\centering
\includegraphics[width=1.0\columnwidth]{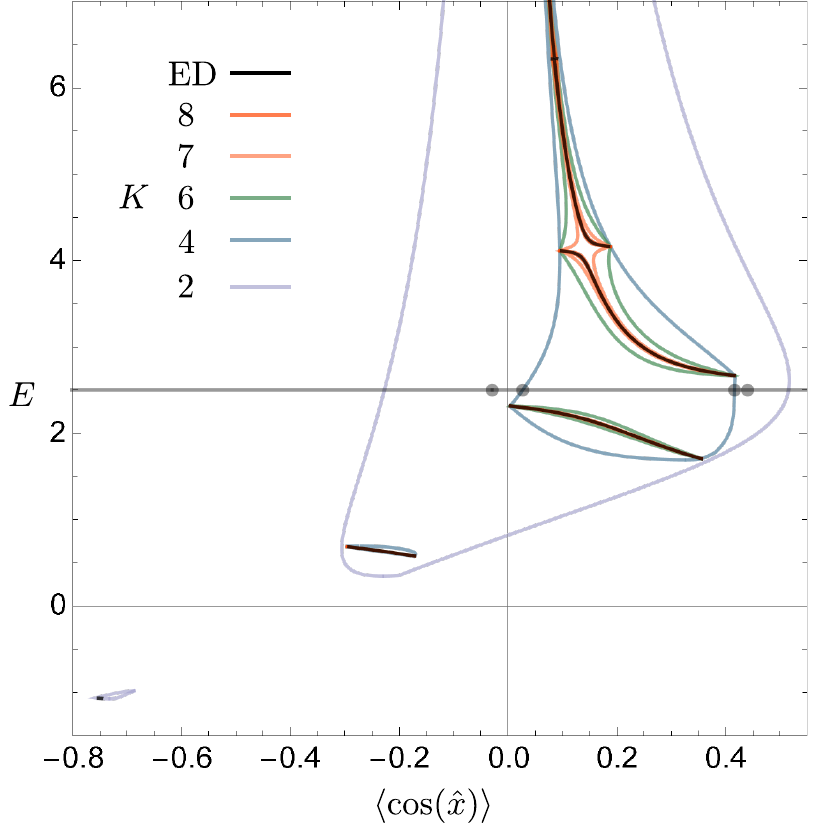}
    \caption{
 Bootstrap solution of Eq.~\eqref{eq:h} for $v = 1$ and an increasing number $K$ of statistical moments, and compared to exact diagonalization (labelled ``ED")~\cite{notebook}. The regions allowed by the bootstrap shrink to six Bloch bands for large values of $K$. Four dots indicate the potential location of the boundary at energy $E=2.5$ and for $K=4$, according to Eq.~(\ref{eq:reduction}).}
    \label{fig:Fig1}
\end{figure}

We apply this technique to compute the spectrum of Eq.~\eqref{eq:h}, as a function of
$\langle \cos(\hat{x}) \rangle$, comparing to results from exact diagonalization for a potential $v=1$, see Fig.~\ref{fig:Fig1}. A similar spectrum was recently obtained in the bootstrap study of gauge invariance of a charged particle on a circle~\cite{aikawa2021application,berenstein2021bootstrapping2}.
The computations are done by increasing the number of constrained statistical moments, when increasing $K$, and for $L = 1$, which implies that the moments only include powers of $\hat{p}$ with $0\leq s\leq 2$. 
We find that including statistical moments of the two conjugate variables is crucial for the convergence of the bootstrap. When we perform the boostrap in position or in momentum only, respectively with moments $\langle \exp(in\hat{x})\rangle$ ($L=0$) or $\langle p^s \rangle$ ($K=0$), we find that some solutions consistent with the bootstrap are absent in exact diagonalization, even for large values of $K$ or $L$ respectively. For $L=1$ and large values of $K$, we see in Fig.~\ref{fig:Fig1} the appearance of six Bloch bands that coincide with exact diagonalization, in black and labeled `ED'.
The four dots on the line of constant energy $E=2.5$ illustrate four of the ten eigenvalues of Eq.~\eqref{eq:reduction} for $K=4$. The two correct eigenvalues that correspond to the boundary are uniquely determined by requiring that all but one eigenvalues $\lambda$ of $\mathcal{M}$ are strictly positive. In this way, on a given domain of the energy $E$, we bound the solutions consistent with the bootstrap by continuous lines.

We observe in Fig.~\ref{fig:Fig1} that the bootstrap converges towards the spectrum in exact diagonalization. For $K = 8$, the average relative error is $\Delta E/E \approx 10^{-3}$ on the considered range of energies, and goes down to $\Delta E/E \approx 10^{-7}$ for $K=12$. These values are for the interaction value $v=1$, and are typically less precise for larger values of $v$. There is also loss in numerical precision when increasing $K$ and it can be necessary to use symbolic calculus or arbitrary precision arithmetics.

\section{Statistical approach to compute arbitrary observables}
\label{Sec:Distrib}
A limitation of the bootstrap is that it does not automatically output some physical quantities, such as the Bloch wavevector -- since $\langle \hat{p}\rangle=0$. From Bloch's theorem, exact eigenstates $|\Psi\rangle$ of $\hat{H}$ in Eq.~(\ref{eq:h}) satisfy $\langle x|\Psi\rangle=\Psi(x)=e^{ikx} \Phi(x)$, with $\Phi(x)$ a $2\pi$ periodic function,
and $k$ the Bloch wavevector. We thus have $e^{i2\pi\hat{p}} |\Psi\rangle = e^{i 2\pi k}
|\Psi\rangle$, so the Bloch wavevector $k$ can be determined from the identity $\langle
e^{i2\pi\hat{p}}\rangle = e^{i 2\pi k}$. The Bloch momentum $k$ is thus found from the
Taylor series
\begin{align}
    e^{i2\pi k} = \langle e^{i2\pi\hat{p}} \rangle = \sum_{s=0}^{\infty} \frac{(2i\pi)^{s}}{s!} \langle \hat{p}^s \rangle,
    \label{eq:bloch}
\end{align}
that depends on the infinite set of moments $\langle \hat{p}^s \rangle$.
The evaluation of this expression from a recursion formula that involves $\langle \hat{p}^s \rangle$ is unstable and is sensitive to the noise generated by numerical errors on higher
moment of $\hat{p}$. Instead, we propose to evaluate $\langle e^{i2\pi\hat{p}} \rangle$ over
the probability distributions $\mathcal{P}(p)$ that reproduce exactly a finite number of 
known moments, $\{\langle \hat{p}^0 \rangle, \langle \hat{p}^1 \rangle, ..., \langle \hat{p}^{K} 
\rangle \}$, as $\langle \hat{p}^s\rangle = \int dp \mathcal{P}(p) p^s$.

In practice, there is an infinite set of probability distributions $\mathcal{P}(p)$ that match the
lowest moments. These probability distributions can be expanded over the so-called canonical
distributions, the probability distributions that are non-zero only for a finite number of discrete
momenta $\{p_0,\ldots,p_{K}\}$.  These momenta are poles of the Stieltjes transform of the
underlying probability distribution, and are found to be solutions of the following $(K+1)$-th order
polynomial equation~\cite{urrea_01,beck_21}:
\begin{align}
h_{K}(p) + \nu h_{K-1}(p) = 0,
\label{eq:nevaanlina}
\end{align}
with $\nu$ a scalar labeling each canonical solution $\mathcal{P}_{\nu}$ and $h_{K}(p)$ a unique set of orthogonal polynomials with respect to the probability distribution $d\mathcal{P}$~\cite{urrea_01} defined as:
\begin{equation}
h_K(p) = \frac{1}{\Delta_K \Delta_{K-1}}
    \left|
    \begin{array}{cccc}
    \big<p^0\big> & \big<p^1\big> & \cdots & \big<p^K\big> \\
    \big<p^1\big> & \big<p^2\big> & \cdots & \big<p^{K+1}\big> \\
    \vdots  & \vdots  & \ddots & \vdots  \\
    \big<p^{K-1}\big> & \big<p^{K}\big> & \cdots & \big<p^{2K-1}\big> \\
    1 & p & \cdots & p^K
    \end{array}
    \right|,
\end{equation}
with $\Delta_K = \mathrm{det} [\big<p^{i+j}\big>]_{i,j=0\ldots K}$.

The probability distribution vanishes for momenta away from the solutions $\{p_0,\ldots,p_{K}\}$ of
Eq.~\eqref{eq:nevaanlina}, so that the canonical probability distributions obey:
\begin{equation}
\mathcal{P}_{\nu}(p) = \sum_{i=0}^{K} \pi_i \delta(p-p_i).
\end{equation}
Both the support $\{p_0,\ldots,p_K\}$ of $\mathcal{P}(p)$ and the amplitudes $\pi_i\equiv\pi(p_i)$ depend on the scalar $\nu$, which parametrizes the set of canonical distributions. The amplitudes $\pi_i$ can be deduced from a bootstrap
condition, similar to the one used to analyze magnetoresistance data in
Refs.~\cite{beck_87,beck_21}. Indeed, a physical probability distribution $\mathcal{P}_{\nu}(p)$
must satisfy the consistency condition that $\mathcal{M}_{\sigma\tau} = \langle \hat{p}^{\sigma + \tau} \rangle$ is positive semi-definite for the $K$ lowest moments $\{\langle \hat{p}^0 \rangle, \langle \hat{p}^1 \rangle, \ldots, \langle \hat{p}^{K} \rangle \}$. We want to assess if this distribution has the amplitude $\pi(p^\star)$ at $p = p^\star$, as described by the Dirac distribution $\pi(p^\star) \delta(p-p^\star)$.
For this purpose, we subtract $\pi(p^\star) \delta(p-p^\star)$ to the target probability
distribution $\mathcal{P}(p)$, defining the new matrix of moments $\mathcal{M}^{\prime}_{\sigma\tau} =
\mathcal{M}_{\sigma\tau} - \pi(p^\star) D_{\sigma\tau}$, where $D_{\sigma\tau} \equiv
(p^\star)^{\sigma+\tau}$. If $\mathcal{M}^{\prime}$ is positive
semi-definite, then we underestimate the amplitude of $\pi(p^\star)$, and if $\mathcal{M}^{\prime}$
is negative, then we overestimate the amplitude of $\pi(p^\star)$. Thus, the smooth envelope
function $\pi(p^\star)$ describes the probability distribution $\mathcal{P}$ when one eigenvalue of
$\mathcal{M}^{\prime}$ cancels, \textit{i.e.}:
\begin{align}
\left( \mathcal{M} - \pi(p^\star) D \right) {\bf w} = 0.
\label{eq:pip0}
\end{align}
Since $D$ is a matrix of rank $1$, it has an unique non-zero eigenvalue $a(p^\star) = \sum_{i = 0}^{K} (p^\star)^{2i}$ with associated eigenvector $v(p^\star) = (1, p^\star, (p^\star)^2, \ldots, (p^\star)^K)$, and we can easily project Eq.~\eqref{eq:pip0} on $v(p^\star)$ and find the unique solution~\cite{beck_87}
\begin{align}
    \pi(p^\star) = \frac{|{\bf v}(p^\star)|^2}{a(p^\star)}\left(\sum_{i=1}^{K} \frac{({\bf e}_i\cdot {\bf v}(p^\star))^2}{\lambda_i}\right)^{-1},
    \label{eq:pistar}
\end{align}
where $(\lambda_i,{\bf e}_i)$ are eigenvalues and eigenvectors of $\mathcal{M}$. Since $\mathcal{M}$ is positive semi-definite, the eigenvalues $(\lambda_i)_{i=1..K}$ are also positive semi-definite. In practice, some eigenvalue are very close to zero and appear negative due to floating point errors. This can lead to numerical instability, that we avoid by evaluating~\eqref{eq:pistar} with an absolute value on $\lambda_i$~\cite{beck_87}.

\begin{figure}[htb]
\centering
\includegraphics[width=1.0\columnwidth]{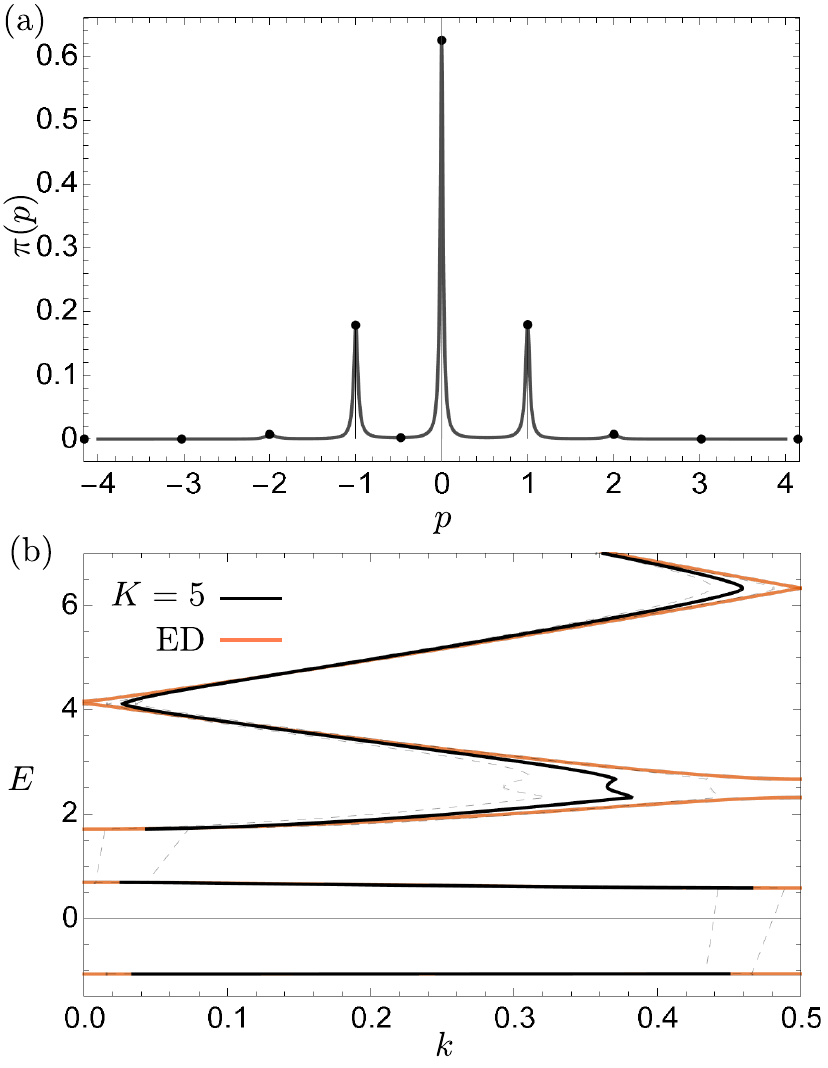}
    \caption{
    Reconstruction of the Bloch momentum for $v = 1$ in the quantum bootstrap with $K = 5$~\cite{notebook}. (a) Envelope function $\pi(p)$ of the canonical probability distribution in the fundamental state at $k = 0$. The dots correspond to the canonical solution for $\nu = 10$ in Eq.~\eqref{eq:nevaanlina}. (b) Bloch band spectrum of the periodic quantum Hamiltonian obtained as the average (in black) of the canonical solutions at $\nu=0$ and $\nu = \infty$ and the corresponding variance (in thin dashed lines). We compare our result to the exact diagonalization, in orange.} 
    \label{fig:Fig2}
\end{figure}

In Fig.~\ref{fig:Fig2}(a) we show the smooth envelope function $\pi(p^\star)$~\eqref{eq:pistar} and
indicate with dots the canonical distribution $\mathcal{P}_{\nu}$ for $\nu =
10$~\eqref{eq:nevaanlina}, for the fundamental state at $k=0$ and a bootstrap with $K = 5$. We did not consider larger values of $K$ because of the considerable increase in the calculation time. We observe that peaks in the probability distribution are approximately equally spaced. This is
reminiscent of the probability distributions associated to eigenstates of~Eq.~\eqref{eq:h}. Indeed,
according to Bloch's theorem, a eigensolution with Bloch wavevector $k$ can be expanded in the form
$\Psi_k(x) = e^{ikx}\sum_{n\in \mathbb{Z}} C_n e^{inx}$ which corresponds to the probability
distribution $\mathcal{P}(p) = \int dx~ |\Psi_k(x)|^2 = \sum_{n\in \mathbb{Z}} |C_n|^2
\delta(n+k-p)$, with poles at momenta $p_n = n + k$ with $n$ an integer. We observe that in the
infinite set of canonical solutions derived from Eq.~\eqref{eq:nevaanlina}, only the solutions at
$\nu = 0$ and $\nu = \infty$ have poles matching the exact solution.

The probability distribution is not uniquely determined from the lowest moments $\{\langle \hat{p}^0 \rangle, \langle \hat{p}^1 \rangle, ..., \langle \hat{p}^{K} \rangle \}$ and is in general a linear combination over all $\nu\in\mathbb{R}$ of the canonical probability distributions $\mathcal{P}_{\nu}$ obtained from Eq.~\eqref{eq:nevaanlina}. Since only $\nu=0$ and $\nu=\infty$ reproduce the correct quantization condition, we restrict the solution to the corresponding branches and estimate the Bloch momentum as $k = \frac12(k_{\nu=0} + k_{\nu=\infty})$ with a precision $\Delta k = \frac12| k_{\nu=0} - k_{\nu=\infty}|$.
We then obtain the Bloch spectrum in Fig.~\ref{fig:Fig2}(b) that associates a Bloch momentum to each energy eigenstate, comparing well with the result of exact diagonalization (in orange). The thick black line show the average Bloch momentum we obtain, while the gray dashed lines help represent the variance $\Delta k$. This reconstruction method does not rely on a particular symmetry but only on bootstrap and could extend to other studies for computing observables that are not easy to access with a recursion~\cite{aikawa2021application,berenstein2021bootstrapping2}. The bootstrap consistency condition is equivalent to the truncated Hamburger moment problem, where one search the probability distributions that are consistent with a finite sequence of moments. We use this connection in our reconstruction technique~\cite{urrea_01} and further investigation of this connection could prove useful for future work on the bootstrap.

\section{Conclusion}
\label{Sec:Conclusion}
In this work, we have explored the bootstrap method for the quantum mechanical problem of a single particle in a cosine potential. While this problem is sufficiently simple for a direct numerical diagonalization, implementation of the bootstrap required the development of several tools that can be useful in more ambitious contexts. For instance, we demonstrate that the dimension of the search space can be reduced by one, thus helping accelerate the calculations. In addition, we propose a probabilistic method to compute observables that are not easily obtained from bootstrap equations. We conclude this article by discussing some perspectives of bootstrap ideas for other situations. In quantum mechanics, periodic potentials that involve a finite number of resonators could be studied by the same method, although the recursion is more complex to set up. It does not seem feasible to extend our study to spatial dimensions above one however, because of the many degrees of freedom involved. It might also be worthwhile to examine whether a matrix model extension of the cosine potential can be solved by bootstrap. Another interesting avenue concerns whether bootstrap could apply to random tight binding lattices, where similar ideas have been developed with a focus on the local environment~\cite{amorphous,Marsal30260}.

{\it Acknowledgements.-} We thank M. Debertolis for discussions. S. T. acknowledge financial support from the European Union Horizon 2020 research and innovation program under grant agreement No829044 (SCHINES).

\appendix
\section{Recursion for statistical moments involving conjugate variables}
\label{App}
We derive first the double recursion Eq.~(\ref{DoubleRecursion}) used to find the mixed moments
of the problem. Starting with $\langle \hat{H} e^{inx} p^{s} \rangle = E \langle e^{inx} p^{s} \rangle$,
we have:
\begin{align}
& \langle \hat{p}^2 e^{in\hat{x}}\hat{p}^{s} + v(e^{i\hat{x}} + e^{-i\hat{x}})e^{in\hat{x}}\hat{p}^{s} \rangle 
= E \langle e^{in\hat{x}} \hat{p}^{s} \rangle.
\end{align}
Commuting $\hat{p}$ through $e^{in\hat{x}}$ gives:
\begin{eqnarray}
&& n^2 \langle e^{in\hat{x}} \hat{p}^{s} \rangle + 2n\langle e^{in\hat{x}}\hat{p}^{s+1} \rangle + 
\langle e^{in\hat{x}}\hat{p}^{s + 2}\rangle \\ 
&& + v\langle (e^{i(n+1)\hat{x}} + e^{i(n-1)\hat{x}}) \hat{p}^{s} \rangle = 
E \langle e^{in\hat{x}} \hat{p}^{s} \rangle,
\end{eqnarray}
which is the wanted result.


Finally, since the bootstrap is based on the mixed operator
$\hat{O} = \sum_{n = 0}^{K} \sum_{s=0}^{L} a_{ns} \hat{p}^s e^{i n \hat{x}}$,
this requires the computation of the following averages:
\begin{align}
\mathcal{M}_{tm,sn} = \langle e^{-im\hat{x}}\hat{p}^{t+s} e^{in\hat{x}} \rangle,
\end{align}
that are not explicitly in the form $\langle e^{in\hat{x}} \hat{p}^{s} \rangle$
obtained from the recursion Eq.~(\ref{DoubleRecursion}). Due to this complication,
we have computed the matrix $\mathcal{M}$ for $0\leq s+t \leq 1$ only.
For $s + t = 0$, one has obviously
$\langle e^{-im\hat{x}} e^{in\hat{x}} \rangle = \langle e^{i(n-m)\hat{x}} \rangle$.
For $s + t = 1$, we get:
\begin{align}
\langle e^{-im\hat{x}} \hat{p} e^{in\hat{x}} \rangle &= n \langle e^{i(n-m)\hat{x}} \rangle +
\langle e^{i(n-m) \hat{x}} \hat{p}\rangle\\
&= \frac{n+m}{2} \langle e^{i(n-m)\hat{x}} \rangle,
\end{align}
using Eq.~\eqref{eq:b1} and using the fact that $\langle \hat{p}\rangle=0$. Finally, for
$s + t = 2$, we obtain by the same method:
\begin{eqnarray}
\langle e^{-im\hat{x}} \hat{p}^2 e^{in\hat{x}} \rangle &=&
\left( nm + E \right)\langle e^{i(n-m)\hat{x}} \rangle\\ 
&& - v\left( \langle e^{i(n-m+1)\hat{x}} \rangle + \langle
e^{i(n-m-1)\hat{x}} \rangle\right).
\nonumber
\end{eqnarray}

\bibliography{bibliography}

\begin{thebibliography}{23}
\bibitem [{notebook()}]{notebook}%
  \BibitemOpen
  \bibinfo {note} {This manuscript is distributed with a Mathematica notebook
  to reproduce all the figures in the text.}\BibitemShut {Stop}%
\end{thebibliography}

\end{document}